\documentclass[nocompress,12pt]{spieman}
\usepackage[]{graphicx}
\usepackage{setspace}
\usepackage{hyperref}
\usepackage{amsmath}
\usepackage{tocloft}
\usepackage{tikz}
\usepackage{array}
\usetikzlibrary{shapes.geometric,arrows}
\usetikzlibrary{shapes.multipart}
\usepackage[OT2,T1]{fontenc}
\DeclareSymbolFont{cyrletters}{OT2}{wncyr}{m}{n}
\DeclareMathSymbol{\Sha}{\mathalpha}{cyrletters}{"58}

\usepackage{xcolor,colortbl}
\definecolor{black}{rgb}{0.3,0.3,0.3}
\definecolor{red}{rgb}{1.0,0.0,0.0}
\definecolor{blue}{rgb}{0.0,0.0,1.0}

\title{Point-spread function ramifications and deconvolution of a signal dependent blur kernel due to interpixel capacitive coupling} 

\author{Kevan Donlon\supscr{a}, Zoran Ninkov\supscr{a}, Stefi Baum\supscr{a,b}}

\affiliation{\supscrsm{a}Rochester Institute of Technology, Chester F. Carlson Center for Imaging Science, 54 Lomb Memorial Drive, Rochester, New York, 14623\\
\supscrsm{b}University of Manitoba, Department of Physics and Astronomy, 66 Chancellors Cir, Winnipeg, MB, Canada}

\cftpagenumbersoff{figure}
\cftpagenumbersoff{table} 
\begin{document} 
\maketitle 

\begin{abstract}
Interpixel capacitance (IPC) is a deterministic electronic coupling that results in a portion of the collected signal incident on one pixel of a hybridized detector array being measured in adjacent pixels. Data collected by light sensitive HgCdTe arrays which exhibit this coupling typically goes uncorrected or is corrected by treating the coupling as a fixed point spread function. Evidence suggests that this IPC coupling is not uniform across different signal and background levels. This variation invalidates assumptions that are key in decoupling techniques such as Wiener Filtering or application of the Lucy-Richardson algorithm. Additionally, the variable IPC results in the point spread function (PSF) depending upon a star's signal level relative to the background level, among other parameters. With an IPC ranging from 0.68\% to 1.45\% over the full well depth of a sensor, as is a reasonable range for the H2RG arrays, the FWHM of the JWSTs NIRCam 405N band is degraded from 2.080 pix (0.132") as expected from the diffraction pattern to 2.186 pix (0.142") when the star is just breaching the sensitivity limit of the system.  For example, When attempting to use a fixed PSF fitting (e.g. assuming the PSF observed from a bright star in the field) to untangle two sources with a flux ratio of 4:1 and a center to center distance of 3 pixels, flux estimation can be off by upwards of 1.5\% with a separation error of 50 millipixels. To deal with this issue an iterative non-stationary method for deconvolution is here proposed, implemented, and evaluated that can account for the signal dependent nature of IPC.
\end{abstract}


{\noindent \footnotesize{\bf Address all correspondence to}: Kevan Donlon, Rochester Institute of Technology, Chester F. Carlson Center for Imaging Science, 54 Lomb Memorial Drive, Rochester, New York, 14623; E-mail: \linkable{KevanADonlon@gmail.com} }

\begin{spacing}{2} 

\section{Introduction}
\label{sect:intro} 
\begin{figure}
\begin{center}
\begin{tabular}{c}
\includegraphics[height=6.0cm]{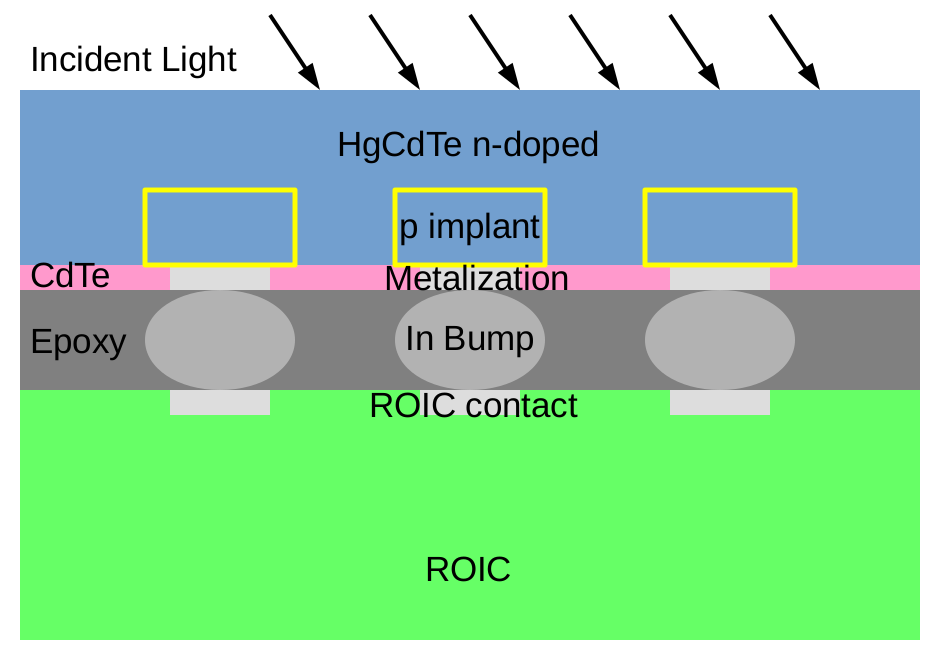}
\end{tabular}
\end{center}
\caption{Cross section illustrating structure of a hybridized imaging array.  Not to scale.}
\label{fig:cartoon_cross}
\end{figure}

Hybridization has become a standard portion of the process for most detectors which utilize unconventional semiconductors for photon detection and even some silicon arrays.  Hybridized detectors are composed of separate photodiode and read-out circuit layers, which are connected to each other using indium bump bonds to form electrical contact as illustrated in figure\ref{fig:cartoon_cross}. In this type of detector array, interpixel capacitance (IPC) is a mechanism by which electronic cross talk occurs between pixels\cite{Moore06}.  Conductive regions in adjacent pixels, responsible for charge collection and storage, are separated by insulating regions, which permit the presence of electric fields.  This configuration gives rise to a classic capacitor\cite{Ohanian07}.  The presence of this capacitance results in a coupled relationship between a pixel's charge and its neighbors' electrostatic potentials.  The readout from a pixel corresponds to this electrostatic potential, yielding the final result that signal collected on one pixel is attributed, not just to that pixel, but also to its neighbors.  This type of cross-talk is distinct from diffusion; diffusion occurs when charge carriers, generated under one pixel, are collected in a neighboring pixel.  In the case of IPC coupling the charge carriers do not move between pixels; they are collected in one pixel and their collection impacts the electrostatic state of nearby pixels.  Early observation of IPC was made due to the coupling's reduction of measured Poissonian noise through the correlation introduced\cite{Moore03}\cite{Moore04}.   By examining the autocorrelation of flat fields the magnitude of the IPC coupling could be assessed\cite{Moore06}\cite{Moore04}.  This correlation invalidated an essential assumption\cite{Moore06} for calculation of a sensor's conversion gain using the photon transfer method\cite{Janesick01}.  This, in turn, necessitated the adoption of a direct capacitive comparison method to accurately determine the conversion gain\cite{Finger06}.  However, the impact that IPC coupling has on collected data remains an issue.  Correction methods for IPC have been proposed\cite{McCullough08} but these methods act to deconvolve a constant coupling.  The underlying semi-conductor physics\cite{Sze07} indicates that this assumption of constant coupling may not be true when a sensor is exposed to spatially varying illumination,  a variable which autocorrelation methods are incapable of exploring.  Simulations from first principle\cite{Donlon17} and measurements obtained using cosmic ray exposures and hot pixels of various intensities\cite{Donlon16}\cite{Cheng09} have verified that the coupling of signal by IPC is a function of the signal level.

As pixel sizes used in modern focal plane arrays continue to decrease, the distance between conductive elements of adjacent pixels is reduced, resulting in an increase of interpixel capacitance\cite{Ohanian07}.  This increase in capacitance results in a greater coupling\cite{Moore06} which has inspired exploration of the impact that IPC may have on photometric and astrometric observations for future missions such as WFIRST\cite{Kannawadi15}\cite{Kannawadi16}.  Until this point, these examinations have not included the signal dependence of IPC.

The goal of this work is to rigorously model the photometric and astrometric effects of a scene dependent IPC as has been predicted\cite{Donlon17} and observed\cite{Donlon16}\cite{Cheng09}.  This approach allows for simultaneous development of an efficient and effective decoupling technique to aid in the restoration of photometric and astrometric accuracy.
\section{Mathematical methods}
A strict mathematical model of the application of IPC coupling allows for examination in a way that lends itself to the development of an iterative approximation technique by which decoupling can be achieved.
\subsection{A rigorous definition of coupling}
The fundamental pixel to pixel issue is that on any particular pixel; $i,j$, the value measured in that pixel; $M(i,j)$, is not the signal collected in that pixel; $S(i,j)$\cite{Moore03}.  A portion of signal collected on a pixel couples capacitively to its neighbors, while simultaneously the neighbors couple capacitively to the initial pixel.  These values are connected by a coupling kernel defined with nearest neighbor coupling as\cite{Moore04}:
\begin{equation}
\label{eq:blur}
K(i,j)=\left(\begin{array}{ccc} 0&\alpha&0\\\alpha&1-4\alpha&\alpha\\0&\alpha&0\end{array}\right)
\end{equation}
by the relation\cite{Moore04}:
\begin{equation}
\label{eq:conv}
M(i,j) = K(i,j) \ast S(i,j)
\end{equation}
This results in a fraction, $\alpha$ referred to as the IPC coupling coefficient\cite{Moore04}, of signal moving from a pixel to each of its neighbors.  Expressing this relationship in the discrete form with the convolution expanded we have the following relationship:
\begin{equation}
\label{eq:expansion}
M(i,j) = S(i,j) + \sum_{m=i-1}^{i+1} \sum_{n=j-1}^{j+1}[\alpha(i,j,m,n) \cdot S(m,n)] - \sum_{m=i-1}^{i+1} \sum_{n=j-1}^{j+1}[\alpha(m,n,i,j) \cdot S(i,j)]
\end{equation}
Where $M(a,b)$ is the value measured on pixel $a,b$; $S(a,b)$ is the signal collected on pixel $a,b$; $\alpha(a,b,c,d)$ is the fraction of signal coupled from pixel $c,d$ onto pixel $a,b$; and $i,j$ and $m,n$ are integer pairs to represent pixel locations.  In this way the readout from each pixel is the signal collected in the initial pixel plus the signal coupled from the neighbors into the initial pixel, minus the signal coupled from the initial pixel into the neighboring pixels.

If however, the coupling coefficient varies as a function of the pixel signal level, then the coupling cannot be expressed strictly as a convolution.  Instead the coupling needs to be applied on a pixel by pixel basis depending on the signal collected in both pixels involved in the coupling; that is to say $\alpha(a,b,c,d) = \alpha(S(a,b),S(c,d))$.  Evidence from both a first principle, semi-conductor physics approach\cite{Donlon17}, and data analysis of coupling from isolated single pixel events\cite{Donlon16}\cite{Cheng09} indicate that the coupling coefficient varies as a function of the pixel signal level.  Therefore, this extended form is a more accurate characterization than a simple convolution.  Furthermore, in the case where the range of alpha is small, this approach converges identically to the case of discrete convolution.

\subsection{Decoupling}
To allow equation\ref{eq:expansion} to be solvable for $S$ we must perform an approximation;  taking $\alpha(S(a,b),S(c,d))<<1.00$ we can make the first order approximation of $\alpha(S(a,b),S(c,d)) \cdot S(c,d) \approx \alpha(M(a,b),M(c,d)) \cdot M(c,d)$.  This allows for the expression of an approximation of $S$, indicated by $\hat{S}$ in terms of $M$:
\begin{equation}
\begin{split}
\label{eq:dec}
\hat{S}(i,j) = &M(i,j) - \sum_{m=i-1}^{i+1} \sum_{n=j-1}^{j+1}\biggl[\alpha\biggl(M(i,j),M(m,n)\biggr) \cdot M(m,n)\biggr] \\& + \sum_{m=i-1}^{i+1} \sum_{n=j-1}^{j+1}\biggl[\alpha\biggl(M(m,n),M(i,j)\biggr) \cdot M(i,j)\biggr]
\end{split}
\end{equation}
This equation tells us that that we can approximate the signal collected in a pixel as the value measured in that pixel minus the signal that would have coupled into the pixel from each neighbor, plus the signal that would have coupled out from the pixel into each neighbor.
Our approximation; $\hat{S}$, is now closer to $S$ than our initial observation; $M$, provided that $\alpha(a,b,c,d)$ is strictly signed.  In this way we can reform our earlier approximation and instead use $\alpha(S(a,b),S(c,d)) \cdot S(c,d) \approx \alpha(\hat{S}(a,b),\hat{S}(c,d))\cdot \hat{S}(c,d)$.  Using this method we can devise an iterative approach, not entirely dissimilar to the Eular method, for evaluating successive approximations for the signal collected, $S_q$ for the qth approximation.
\begin{equation}
\begin{split}
\label{eq:full_dec}
\hat{S}_{q}(i,j) = &M(i,j) - \sum_{m=i-1}^{i+1} \sum_{n=j-1}^{j+1} \biggl[ \alpha\biggl(\hat{S}_{q-1}(i,j),\hat{S}_{q-1}(m,n)\biggr) \cdot \hat{S}_{q-1}(m,n)\biggr]  \\& + \sum_{m=i-1}^{i+1} \sum_{n=j-1}^{j+1}\biggl[\alpha\biggl(\hat{S}_{q-1}(m,n),\hat{S}_{q-1}(i,j)\biggr) \cdot \hat{S}_{q-1}(i,j)\biggr]
\end{split}
\end{equation}
\begin{equation}
\label{eq:first_guess}
\hat{S}_{0}(a,b)=M(a,b)
\end{equation}

This process looks at the output frame, calculates what the coupling to and from every pixel would have been if that were the input frame, and then corrects each pixel by that difference.  It then takes that 1st guess at correction as the input frame, calculates what the coupling to and from every pixel would have been in that case, then corrects the measured frame by that difference.  This process continues until the pixel by pixel difference between successive estimates of the input frame approaches zero.

\subsection{Example calculation}
Consider a small 4x4 array with a signal incident slightly off center in such a way that the following signals are recorded on a focal plane array:

\begin{center}
\begin{tabular}{|m{1.5cm}|m{1.5cm}|m{1.5cm}|m{1.5cm}|}
\hline
\cellcolor{blue!45}0.00 & \cellcolor{blue!45}0.00 & \cellcolor{blue!45}0.00 & \cellcolor{blue!45}0.00 \\
\hline
\cellcolor{blue!45}0.00 & \cellcolor{blue!45}2000.00 & \cellcolor{blue!45}4000.00 & \cellcolor{blue!45}0.00 \\
\hline
\cellcolor{blue!45}0.00 & \cellcolor{blue!45}4000.00 & \cellcolor{blue!45}10000.00 & \cellcolor{blue!45}0.00 \\
\hline
\cellcolor{blue!45}0.00 & \cellcolor{blue!45}0.00 & \cellcolor{blue!45}0.00 & \cellcolor{blue!45}0.00 \\
\hline
\end{tabular}
\end{center}

Due to the range of values contained in this array, the coupling will vary between pixel pairs.  We will take the coupling to be governed by the following equation, where $S$ it the signal incident on a pixel and $N$ is the signal incident on its neighbor:

\begin{equation}
\alpha(S,N) = 0.4 \cdot exp(-\frac{|S-N|}{20000.0}) + 0.4 \cdot exp(-\frac{(S^2+N^2)^{\frac{1}{2}}}{10000.0 \cdot \sqrt{2}}) + 0.65 [\%]
\label{eq:functional_alpha}
\end{equation}

\begin{figure}
\begin{center}
\begin{tabular}{c}
\includegraphics[height=6.0cm]{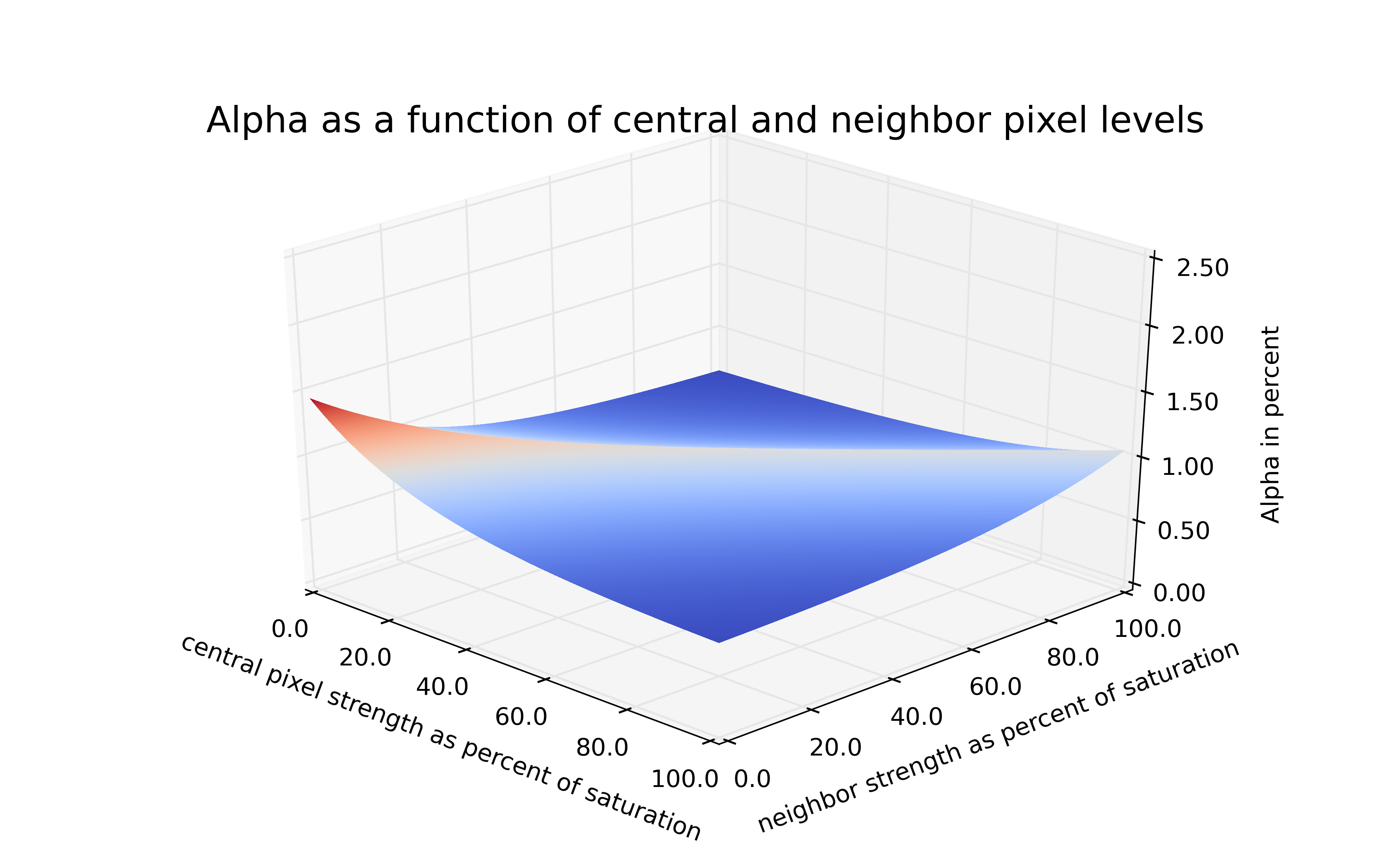}
\end{tabular}
\end{center}
\caption{3-dimensional plot of the coupling coefficient as a function of signal and neighbor strength.}
\label{fig:2.3.1}
\end{figure}

\begin{figure}
\begin{center}
\begin{tabular}{c}
\includegraphics[height=6.0cm]{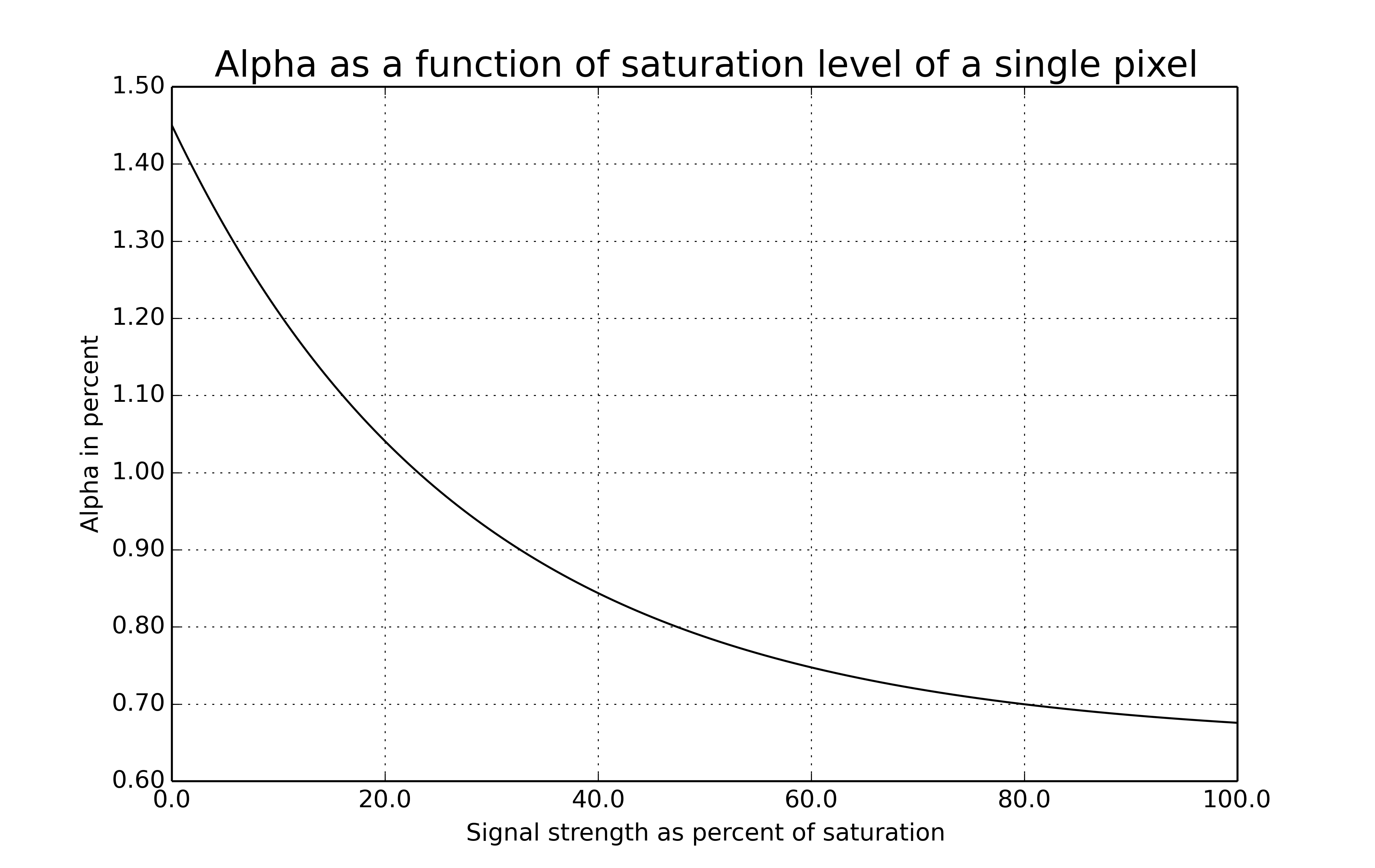}
\end{tabular}
\end{center}
\caption{Variation of $\alpha$ with a zero magnitude neighbor strength.  This plots equation \ref{eq:functional_alpha} with N=0.}
\label{fig:2.3.2}
\end{figure}

This form has the best case behavior of 0.68\% coupling when observing a bright point source over a dim background and increases to a coupling of 1.45\% when observing a dim point source over a background of similar strength.  This form is based on analysis of single pixel events\cite{Donlon16}\cite{Cheng09} combined with simulation results exploring the impact of neighbor pixel strength\cite{Donlon17}.  In addition, it meets the criteria that $\alpha(a,b)=\alpha(b,a)$.  Though it may not perfectly reflect the behavior of IPC in physical sensors, it trends in the directions and of the magnitudes informed by prior data and first principle simulations.  This equation is indicated by a surface plot in figure\ref{fig:2.3.1} and by its zero background cross section in figure\ref{fig:2.3.2}.

\begin{figure}
\begin{center}
\begin{tabular}{c}
\includegraphics[height=12.0cm]{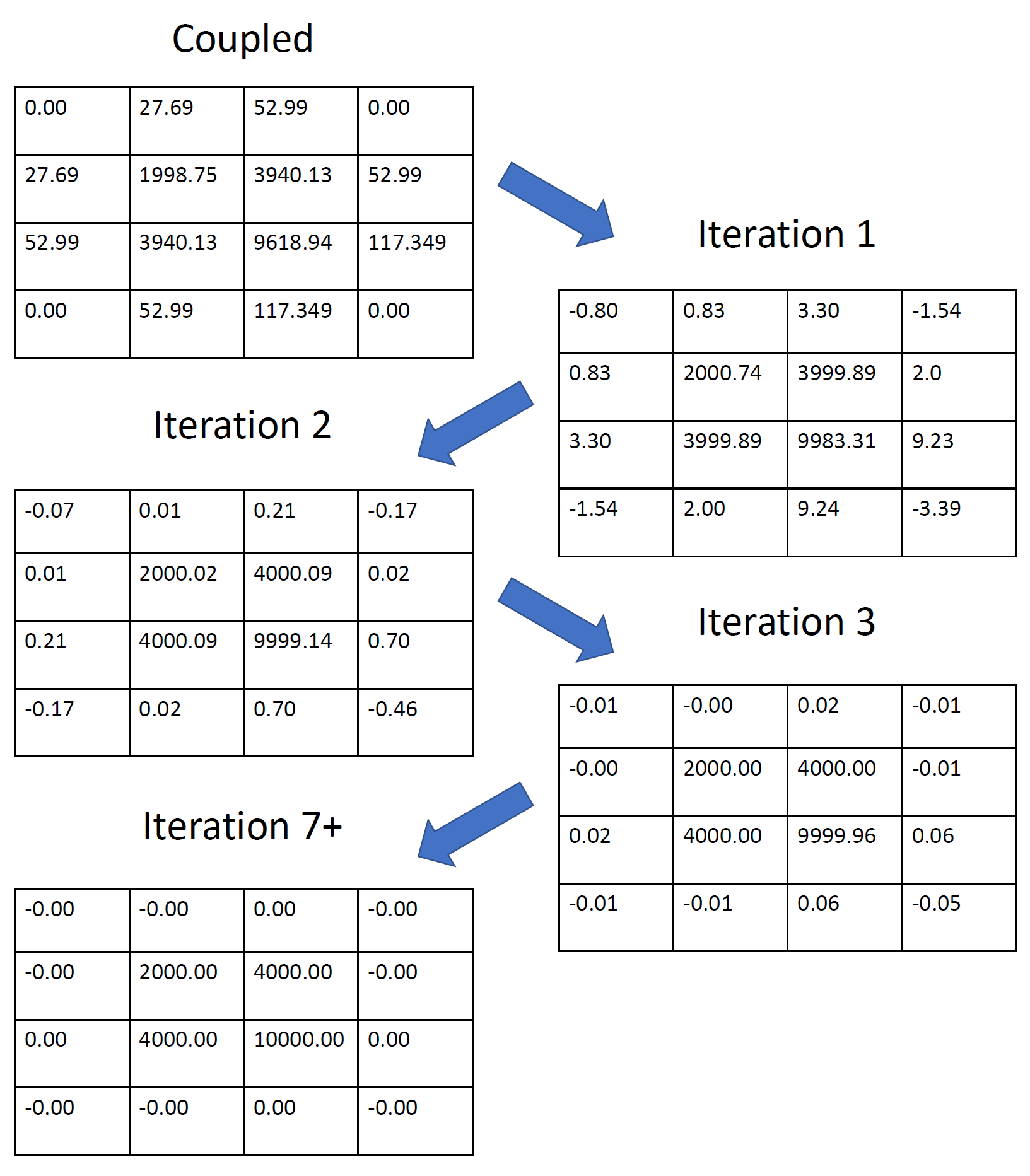}
\end{tabular}
\end{center}
\caption{The result of the described iterative deconvolution algorithm on the array coupled and presented in table\ref{table:coupling} after 0, 1, 2, 3, or 7+ iterations}
\label{fig:dec_flow}
\end{figure}

\clearpage
In this case, the couplings would appear as illustrated in the following\ref{table:coupling}:
\begin{center}
\begin{tabular}{|m{1.5cm}|m{1.5cm}|m{1.5cm}|m{1.5cm}|m{1.5cm}|m{1.5cm}|m{1.5cm}|}
\hline
\cellcolor{blue!45}0.00 & \cellcolor{red!33}1.45\% &\cellcolor{blue!45}0.00 & \cellcolor{red!33}1.45\% &\cellcolor{blue!45}0.00 & \cellcolor{red!33}1.45\% &\cellcolor{blue!45}0.00 \\ 
\hline
\cellcolor{red!33}1.45\%&\cellcolor{black!95}&\cellcolor{red!33}1.39\%&\cellcolor{black!95}&\cellcolor{red!33}1.33\%&\cellcolor{black!95}&\cellcolor{red!33}1.45\% \\ 
\hline
\cellcolor{blue!45}0.00 & \cellcolor{red!33}1.39\%&\cellcolor{blue!45}2000.00 &\cellcolor{red!33}1.35\%&\cellcolor{blue!45}4000.00 & \cellcolor{red!33}1.33\% &\cellcolor{blue!45}0.00 \\ 
\hline
\cellcolor{red!33}1.45\%&\cellcolor{black!95}&\cellcolor{red!33}1.35\%&\cellcolor{black!85}&\cellcolor{red!33}1.22\%&\cellcolor{black!95}&\cellcolor{red!33}1.45\% \\ 
\hline
\cellcolor{blue!45}0.00 &\cellcolor{red!33}1.33\%&\cellcolor{blue!45}4000.00 &\cellcolor{red!33}1.22\%&\cellcolor{blue!45}10000.00 &\cellcolor{red!33}1.17\% &\cellcolor{blue!45}0.00 \\ 
\hline
\cellcolor{red!33}1.45\%&\cellcolor{black!95}&\cellcolor{red!33}1.33\%&\cellcolor{black!95}&\cellcolor{red!33}1.17\%&\cellcolor{black!95}&\cellcolor{red!33}1.45\%\\ 
\hline
\cellcolor{blue!45}0.00 &\cellcolor{red!33} 1.45\%&\cellcolor{blue!45}0.00 & \cellcolor{red!33}1.45\%&\cellcolor{blue!45}0.00 & \cellcolor{red!33}1.45\%&\cellcolor{blue!45}0.00 \\ 
\hline
\end{tabular}
\end{center}
\begin{table}[H]
\caption{A table containing pixel values in the dark blue and the coupling experienced between each pair of neighbors in light red.  Each pixel has four associated couplings, one with each of its neighbors, which is a function of the pixel and neighboring pixel levels.}
\label{table:coupling}
\end{table}
Note that the application of a signal dependent IPC results in a weaker fractional coupling in the neighborhood of the brighter pixels and a stronger fractional coupling between the weaker pixels\cite{Donlon17}.  In fact, the strongest fractional coupling is between the pixels of identical value.  When this IPC is applied to this array it yields the readout presented in figure\ref{fig:dec_flow}.

As illustrated in figure\ref{fig:dec_flow} after IPC coupling there is an underestimate of peak pixel intensity by 3.81\% (i.e. 10000 to 9619).  Application of a single iteration of the non-linear iterative deconvolution algorithm will yield the a reduction of error at the peak intensity pixel to 0.1\%(i.e. restoration from 9619 to 9983).  However, not every individual pixel value has moved closer to the true value; the corner values of this array were unchanged by the initial coupling but have now had their values changed.  In fact, these locations now return nonsensical negative values.  These negative values are a result of the approximation of $\alpha \cdot S = \alpha \cdot \hat{S_q}$  not being strictly true.  This error is corrected in higher order iterations as the approximation becomes progressively more accurate.  Application of another iteration yields an error on the peak intensity of 0.009\%.  Fractional error on the peak after three iterations has been reduced from 3.81\% down to 0.0004\% with the maximum error magnitude down to 0.06 values from 381 values.  With seven or more iterations, the error on each pixel present is <0.005 values. These fractional values exist only in the mathematical abstraction to illustrate the precision of this deconvolution algorithm.  For physical sensors, the original 'coupled' readout from figure \ref{fig:dec_flow} would be quantized at some level of precision and the end result could be re-quantized to the same precision.

\subsection{Implementation}
\label{sec:imp}
This deconvolution technique can be considered as comparable to other filtering techniques such as Wiener filtering\cite{Wiener49}, or Lucy-Richardson deconvolution\cite{Lucy74}\cite{Richardson72}.  While each of these deconvolution techniques works to restore an image blurred by a known point-spread function (PSF) they are all unable to adapt in the presence of a well characterized but variable PSF;  they cannot handle a coupling coefficient that changes across a scene.  However, they all have the advantage of being computable in linearithmic time\cite{Easton10} ($O(n log n)$) where as the method described above is constrained to run in quadratic time ($O(n^{2})$).  This is due to the fact that conventional filtering algorithms can exploit the properties of convolution in the Fourier domain to operate on the entire image simultaneously\cite{Easton10}.  The technique described above is constrained to operate in the image domain and must operate pixel by pixel.  However, within an iteration it requires only computationally cheap look-up, multiplication, and addition operations, while not requiring any reference to the newly updated values until the next iteration begins.  As a result, this algorithm is an excellent candidate for parallel implementation.  Though this algorithm converges quickly as illustrated in section 2.3, a convergence constraint is still necessary for implementation. The convergence condition introduced permits the algorithm to either, cycle 20 times, or cycle until the maximum difference between each individual pixel pair in two successive iterations is less than 0.001\% of the greatest pixel magnitude returned in that iteration.  This constraint is typically met after three to five iterations for $\alpha$ peaking on the order of a few percent.

The implementation used in this work involves a C++ wrapper over CUDA\textsuperscript{\textregistered} code on an NVIDIA\textsuperscript{\textregistered} GTX 960M GPU.  Use of this graphics chip did not come without limitations.  The maximum array size for a single operation is limited to 512 by 512.  Additionally, when utilizing double precision floating point values the total array size is limited to 1024 by 1024.  When instead using single precision floating points the array size is permitted to go to 6144 by 6144. However, beyond 4096 by 4096 arrays the write time for data output becomes excessive.  These constraints apply only to this particular implementation on this particular hardware.  A system with arbitrarily large VRAM and arbitrarily many registers could in principle act on an arbitrarily large image in a single operation per GPU core per iteration.

In order to work around the issue of the limited number GPU registers provided by the GTX 960M graphics chip, an image beyond 512 by 512 is partitioned into smaller 512 by 512 subarrays with longer, narrower subarrays to cover the interfaces between them.  For example, a 1024 by 1024 array would be processed as four 512 by 512 subarrays, one 1024 by 40 array to cover the horizontal interfacing region and one 40 by 1024 array to cover the vertical interface.  For a 2048 by 2048 array, the process would include sixteen 512 by 512 subarrays, three 2048 by 40 horizontal subarrays, and three 40 by 2048 vertical subarrays.  It is this partitioning technique that gives the hard limit to the implementation used here of 6144 by 6144.  This implementation requires that any array larger than 512 by 512 be both square, and have a side length divisible by 512.  Beyond 6144 by 6144, the arrays designed to cover the interface regions between the square subarrays are now past the memory limit of the GTX 960M.

Without a GPU implementation the computation time for even smaller images becomes prohibitively long.  A naive serial python CPU based implementation requires on the order of half an hour to run over a 512 by 512 image on a typical laptop computer.  An implementation using GNU Parallel for CPU based parallel processing can drop this to approximately 8 minutes on a multicore processor.  The GPU implementation can run on a 512 by 512 array in approximately two seconds.  Including image read and write support functions, a 512 by 512 array can be fully decoupled in under 10 seconds.  A 2048 by 2048 array can be decoupled with a total run-time of approximately 6 minutes with >80\% of that time being spent on writing out the array as a *.csv file.  Sample code for both a naive python implementation and the full GPU parallel implementation will be available at \url{https://github.com/Donlok/Decouple}. 

\section{IPC and the Point Spread Function}
The point spread function (PSF) of an imaging system indicates how the system as a whole will respond to an incident point source.  For a digital system, it is the final input to output mapping of the optics, sensor response, and digitization of the readout.  A point source incident on the imaging system will be altered by optical diffraction, the pixel response function, discrete sampling, and any crosstalk present in the array.  The end result is a unit volume mapping of a point in scene space into read-out space.

To build a model of the behavior of an imaging system in the presence of IPC a full mathematical model of imaging system has to be constructed first.  To begin this model we start with a signal;  for a single star we can approximate this as a Dirac-delta function\cite{Easton10}.  This function is defined to be an ideal point source; it has zero value over its full domain except at its origin and when integrated over all space, has unit area\cite{Easton10}. This function is convolved by the diffraction pattern of the light;  for a simple lens system we can approximate this as an airy disk or radial $sinc^2$ function.  For the systems considered later we will use a custom PSF calculated from optics of the JWST.  This result is then sampled by the sensor; a step that is mathematicized as multiplication by a comb function\cite{Easton10}.  Instead of being a continuous function, it is now a discrete set of values that can be read out from the sensor one by one.  In the absence of IPC this is the mathematical point where noise is introduced by sampling using a Poissonian distribution to represent photon noise and then adding in samples from a Gaussian distribution to represent the read noise.  In the presence of IPC this process is slightly more complicated.  The Poissonian sampling occurs as normal to represent the shot noise but it is at this point that IPC is applied\cite{Donlon17}.  When IPC is taken as a constant coupling this is done through convolution with a blur kernel\cite{Moore06}\cite{Moore04}.  In this work we allow IPC to vary as a function of signal strength therefore a discrete sum as described in equation\ref{eq:expansion} is required instead.  After this operation occurs we introduce a read noise distribution by addition of a sampled normal distribution.  So the final mathematical form for an individual sampling would be:
\begin{equation}
\label{eq:PSF_full_form}
M(i,j) = \biggl[\biggl(Poisson\Bigl[x,y;\Bigl( S(x,y) \ast Diff(x,y) \Bigr) \cdot \Sha_{pitch}(x,y) \Bigr]\biggr) \ast K(i,j)\biggr] +Normal(i,j;0,\sigma_{read})
\end{equation}
Where $Poisson(x,y;\gamma)$ is a sample from the Poisson distribution with parameter $\gamma$ at location $x,y$; $Diff(x,y)$ is the diffraction pattern in two dimensions; $\Sha_{pitch}(x,y)$ is the Dirac comb in two dimensions with $x$ and $y$ frequency given by the pixel pitch;  and $Normal(i,j;\mu ,\sigma)$ is a sample from the normal distribution with mean of $\mu$ and variance of $\sigma ^2$.
The PSF that will be discussed here is built in the absence of these noise distributions and using $S(x,y)=\delta(x,y)$.  It can be considered as the average resulting from an ensemble of measurements.

This leaves the PSFs reported here defined as follows:
\begin{equation}
\label{eq:IPC_PSF}
PSF(i,j)=\biggl[\biggl(\delta(x,y)\ast Diff(x,y)\biggr) \cdot \Sha_{pitch}(x,y)\biggr]\ast K(i,j)
\end{equation}

These will be compared to what this PSF would have looked like in the absence of IPC:
\begin{equation}
\label{eq:IPC_noPSF}
PSF(i,j)=\biggl(\delta(x,y)\ast Diff(x,y)\biggr) \cdot \Sha_{pitch}(x,y)
\end{equation}

For the analysis provided here $Diff$ is taken as the WebbPSF F405N as provided from WebbPSF revision V available at \url{www.stsci.edu/~mperrin/software/psf_library/} which is provided oversampled 4x allowing offsets in quarter pixel intervals in each dimension\cite{Perrin14}.  This is the diffraction pattern as would be incident onto the James Webb Space Telescope's long wave NIRcam sensor after passing through the imaging optics and the narrow band 4.05$\mu$m filter.  To provide a continuous PSF the figures here presented are sampled at quarter pixel intervals and then interpolated using a cubic spline method.

\subsection{Testing paradigm}
\tikzstyle{frame}=[rectangle,rounded corners,minimum width=3cm,minimum height=1cm,text centered,draw=black,fill=red!30]
\tikzstyle{op}=[trapezium,trapezium left angle=70,trapezium right angle=110,minimum width=2cm,minimum height=1cm,text centered,draw=black,fill=blue!30]
\tikzstyle{end}=[rectangle,rounded corners,minimum width=3cm, minimum height=1cm,text centered,draw=black,fill=green!30]
\tikzstyle{add}=[circle,text centered,draw=black,fill=white]
\tikzstyle{arrow}=[thick,->,>=stealth]
\tikzstyle{line}=[thick,-]
\begin{center}
\begin{tikzpicture}[node distance=1cm, every text node part/.style={align=center}]
	\node(so) [frame,xshift=-2.0cm,yshift= 00.0cm]{Sources};
	\node(bk) [frame,xshift= 2.0cm,yshift= 00.0cm]{Background};
	\node(rn) [frame,xshift= 8.0cm,yshift=-06.0cm]{Read Noise Instance};
	\node(fm) [frame,xshift= 0.0cm,yshift=-03.0cm]{Flux Map};
	\node(ds) [frame,xshift= 4.0cm,yshift=-03.0cm]{Dark Signal};
	\node(si) [frame,xshift= 2.0cm,yshift=-06.0cm]{Simulated Scene};
	\node(csi)[frame,xshift=-3.0cm,yshift=-09.0cm]{Coupled Scene};
	\node(rn2)[frame,xshift= 2.0cm,yshift=-09.0cm]{Read Noise Instance};
	\node(ad1)[add,  xshift= 0.0cm,yshift= 00.0cm]{+};
	\node(ad2)[add,  xshift= 2.0cm,yshift=-03.0cm]{+};
	\node(ad3)[add,  xshift= 4.8cm,yshift=-06.0cm]{+};
	\node(ad4)[add,  xshift=-0.7cm,yshift=-09.0cm]{+};
	\node(css)[end,  xshift=-0.7cm,yshift=-10.5cm]{Coupled Image};
	\node(ss) [end,  xshift= 4.8cm,yshift=-12.0cm]{Truth Image};
	\node(dss)[end,  xshift=-0.7cm,yshift=-13.5cm]{Decoupled Image};
	\draw[line](so)--(ad1);
	\draw[line](ad1)--(bk);
	\draw[arrow](ad1)--(fm);
	\draw[line](fm)--(ad2);
	\draw[line](ad2)--(ds);
	\draw[arrow](ad2)--(si);
	\draw[arrow](si)--(csi);
	\draw[line](si)--(ad3);
	\draw[line](ad3)--(rn);
	\draw[line](csi)--(ad4);
	\draw[line](ad4)--(rn2);
	\draw[arrow](ad4)--(css);
	\draw[arrow](ad3)--(ss);
	\draw[arrow](css)--(dss);
	\node(nor)[op,xshift=8.0cm,yshift=-4.5cm,]{Take Gaussian sample: \\[-0.5cm] $\mu = 0, \sigma=\sigma_{RMS}$};
	\node(op1)[op,xshift=0.0cm,yshift=-1.5cm]{Convolve with $Diff$};
	\node(poi)[op,xshift=2.0cm,yshift=-4.5cm]{Add Poisson noise};
	\node(ipc)[op,xshift=-0.5cm,yshift=-7.5cm]{Apply IPC};
	\node(dec)[op,xshift=-0.7cm,yshift=-12.0cm]{Decouple};
	\draw[arrow](nor)--(rn);
\end{tikzpicture}
\end{center}
\begin{figure}[H]
\caption{A flow chart that outlines the method used to both generate sample coupled images, then deconvolve them in a way that allows for meaningful pixel by pixel assessment. The rounded rectangles represent data arrays, the trapezoids and circles represent mathematical operations applied to those arrays (e.g. convolve with $Diff$), or used to generate those arrays (e.g. take Gaussian sample.)}
\label{tikx:flow_chart}
\end{figure}
In order to assess the accuracy of this decoupling technique the coupling must be applied to known scenes for analysis.  To accomplish this goal the following method was used:
\begin{itemize}
	\item First, a scene image was generated.  The particular scenes examined here include; point sources convolved with the WebbPSF provided, and random levels assigned to each pixel.
	\item Second, a copy of this scene image is produced.  This copy undergoes IPC coupling by examining the values of each pixel and using the signal dependent $\alpha$ defined by equation\ref{eq:functional_alpha} to determine the IPC coupled image.
	\item Third, a read noise distribution was generated by taking uncorrelated samples from a zero mean normal distribution with variance of $\sigma_{r}^2$ for each pixel.  This distribution is added to the original scene image and the coupled copy. These images are referred to as the truth image or $true(i,j)$ and the coupled image or $coupled(i,j)$ respectively.
	\item Fourth, the coupled image is run through the deconvolution algorithm described in equation\ref{eq:full_dec} and using equation\ref{eq:functional_alpha} as the reference coupling.  This output is referred to as the decoupled image or $decoupled(i,j)$.
\end{itemize}
This method is outlined in the flowchart presented in figure\ref{tikx:flow_chart} above.

This computational method simultaneously gives the results that would be expected if IPC had been present, the results in the presence of IPC, and the results after the removal of IPC using the iterative non-linear method described earlier.  Additionally, it underscores the most significant issue with this particular iterative algorithm;  the read noise distribution that is applied to the scene is not a part of the image prior to coupling, but after being introduced, still undergoes decoupling.  This forces a type of trade-space; this technique does not fully uncorrelate neighboring pixels.  Instead it uncorrelates the Poissonian noise and restores the signal accuracy while forcing a correlation on the read noise.  In the absence of read noise this algorithm, both on average and in every individual pixel, restores levels to the uncoupled values.  In the presence of zero mean read noise, this algorithm restores accuracy on average, but any individual pixel has error proportional to the read noise as will be illustrated in the following section.

\subsection{Comparison to Lucy-Richardson deconvolution}

\begin{figure}
\begin{center}
\begin{tabular}{c}
\includegraphics[height=6.0cm]{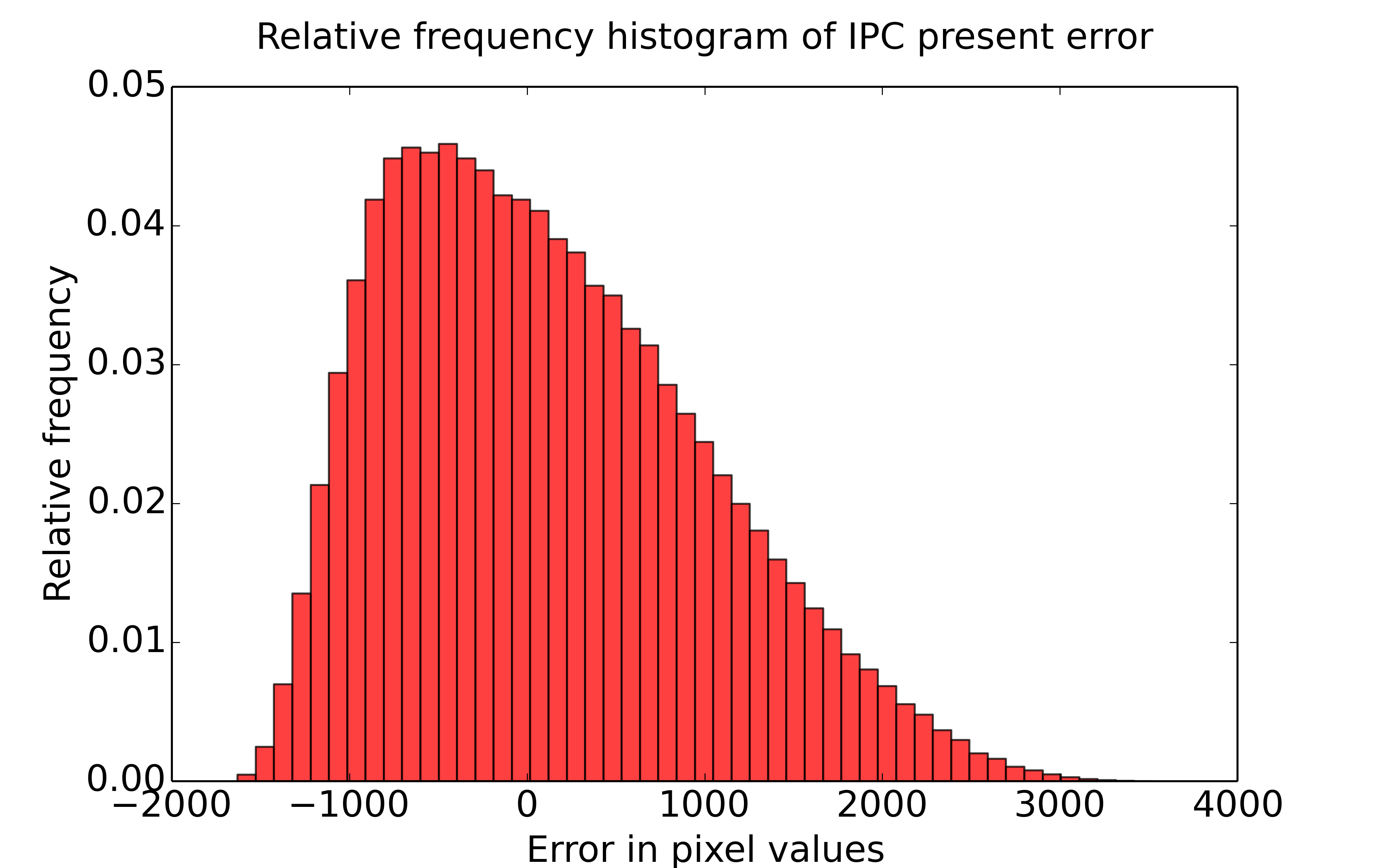}
\end{tabular}
\end{center}
\caption{Histogram of pixel errors for the IPC present case when read noise is set to zero and the input value of each pixel is set to a random value sampled from a uniform distribution from 0 to 60,000.  This histogram contains the results of a pixel by pixel difference, $coupled(i,j)-true(i,j)$.  The maximum pixel error due to IPC is approximately a 3,000 value overestimate}
\label{fig:3.2.2}
\end{figure}

\begin{figure}
\begin{center}
\begin{tabular}{c}
\includegraphics[height=6.0cm]{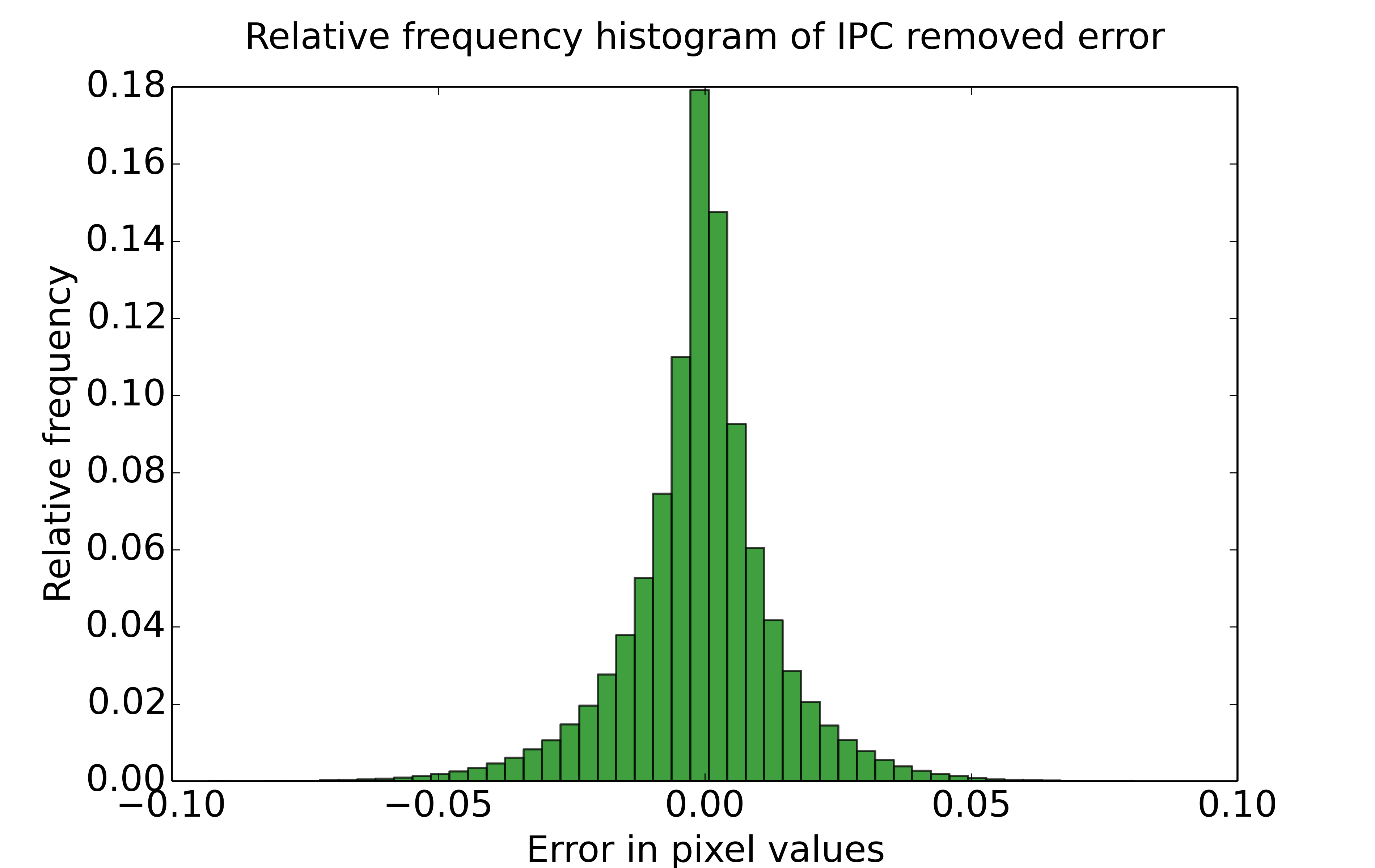}
\end{tabular}
\end{center}
\caption{This histogram contains the results of a pixel by pixel difference, $decoupled(i,j)-true(i,j)$ where the decoupling algorithm used was the iterative method described by\ref{eq:full_dec} after meeting the convergence condition described in section 2.4.  The maximum pixel error after non-linear correction is approximately 0.075 value over or underestimate.}
\label{fig:3.2.1}
\end{figure}

\begin{figure}
\begin{center}
\begin{tabular}{c}
\includegraphics[height=6.0cm]{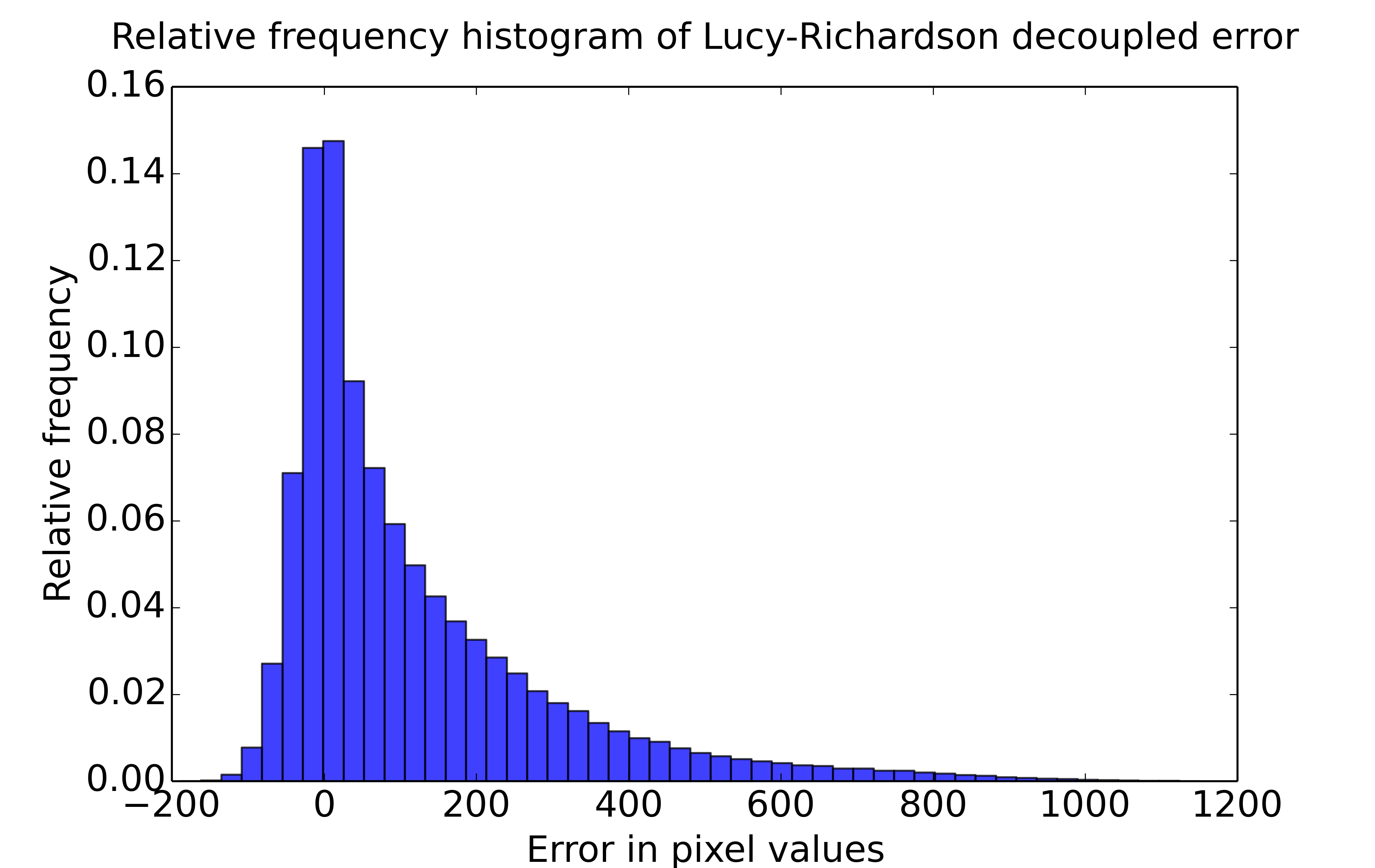}
\end{tabular}
\end{center}
\caption{This histogram contains the results of a pixel by pixel difference, $decoupled(i,j)-true(i,j)$ where the decoupling algorithm used was LR deconvolution.  The maximum pixel error after LR correction is approximately a 1000 value overestimate}
\label{fig:3.2.3}
\end{figure}

In order to evaluate the success of this deconvolution technique it has been compared to established techniques.  The Lucy-Richardson (LR) deconvolution method was selected as it was both the best performing of the standard deconvolution techniques, as well due to the similarities that it has to the method presented here;  both are iterative, though the LR algorithm uses successive iterations to minimize the impact of noise\cite{Lucy74}\cite{Richardson72} rather than as a series of progressively more accurate estimates.  To evaluate the success of each algorithm in restoring an image frame where each pixel was set to a random value sampled from a uniform distribution with range from 0 to 60,000 was used as the input.  A pixel to pixel comparison between the truth image and each of: the LR decoupled image, the iterative non-linear decoupled image, and the IPC coupled image.  Histograms of the pixel to pixel error in each case when no read noise is present are illustrated for, the uncorrected IPC present case in figure\ref{fig:3.2.2} , the iterative method of described here by equation\ref{eq:full_dec} in figure\ref{fig:3.2.1}, and the LR corrected case in figure\ref{fig:3.2.3}.

It is clear in this case that, though the LR deconvolution does reduce error and has a similar average error, the pixel to pixel error as well as the mean absolute error can still be quite large;  even though the error is corrected on average, each pixel can still be a severe over or under estimate.  The iterative non-linear decoupling reduces pixel to pixel error to less than a single count, not just on average, but in every pixel individually.

\begin{figure}
\begin{center} 
\begin{tabular}{c}
\includegraphics[height=6.0cm]{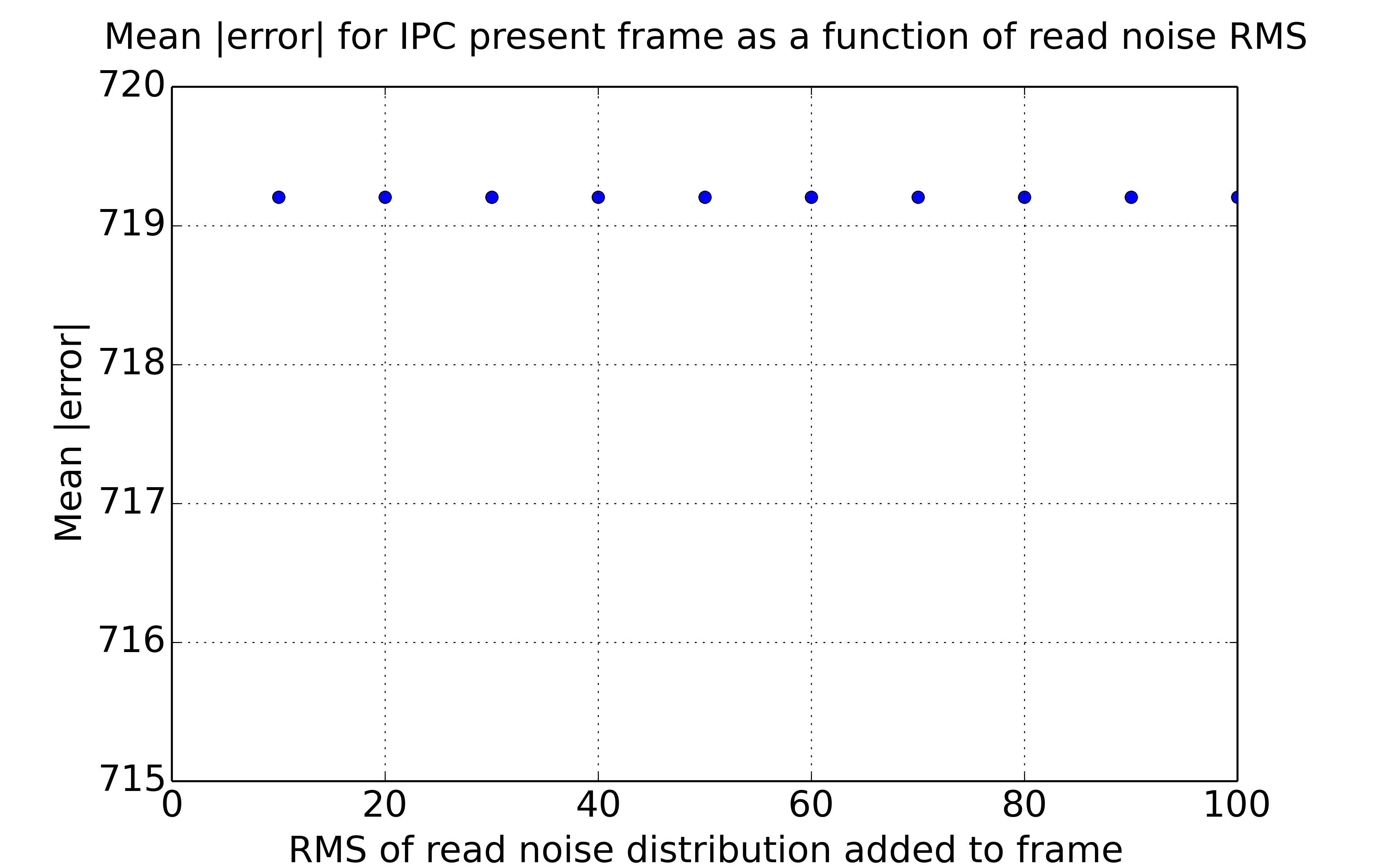}
\end{tabular}
\end{center}
\caption{Mean absolute error as a function of read noise for the IPC present case.  This is indicates that for the random input frame described with $\alpha$ given by equation \ref{eq:functional_alpha} the average magnitude difference between the truth image and the coupled image is ~719 values (i.e. $<|coupled(i,j)-true(i,j)|>\approx719$).  As expected because the read noise is zero mean, this mean absolute error has no read noise dependence.}
\label{fig:3.2.5}
\end{figure}

\begin{figure}
\begin{center}
\begin{tabular}{c}
\includegraphics[height=6.0cm]{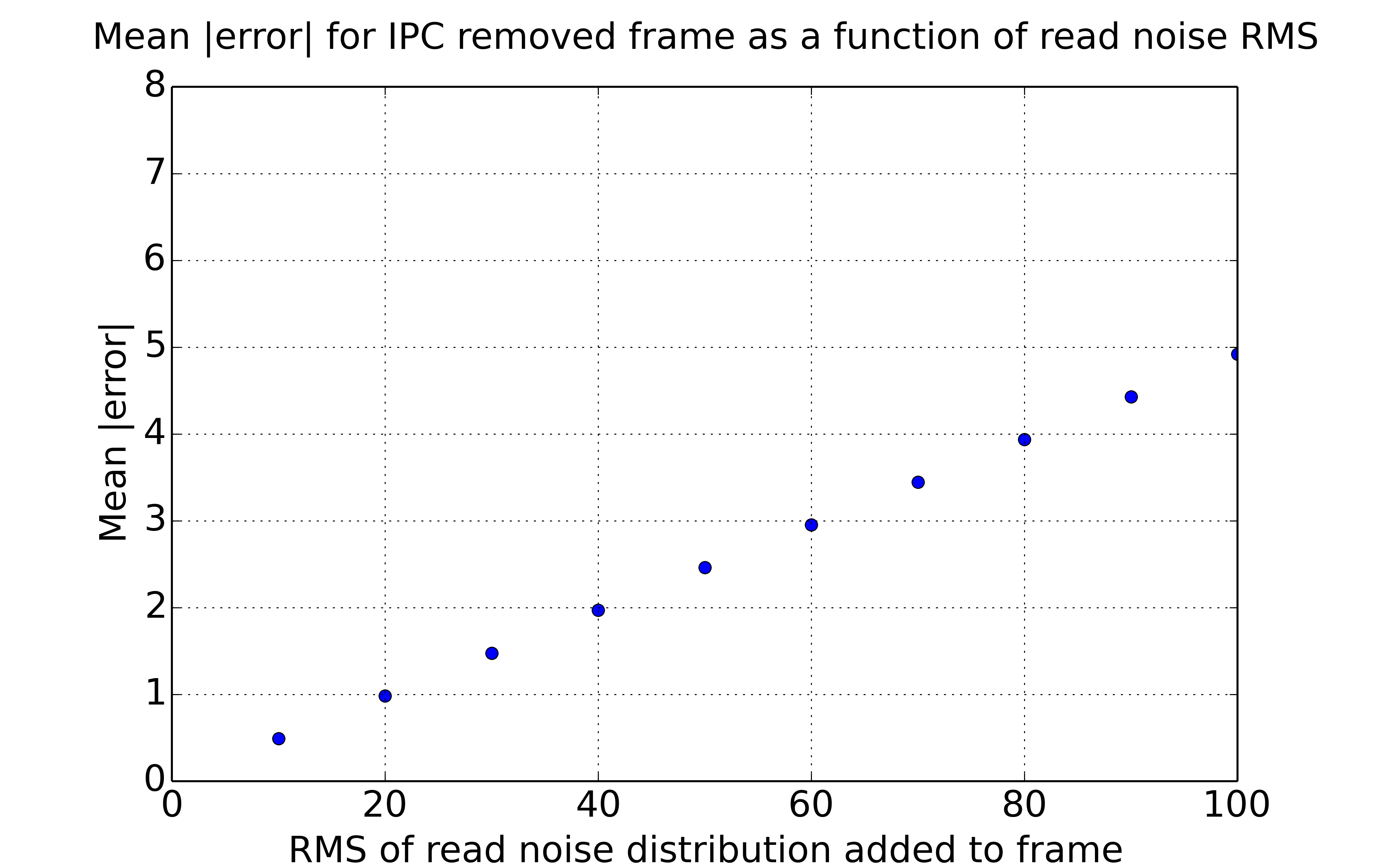}
\end{tabular}
\end{center}
\caption{Mean absolute error as a function of read noise for the non-linear iterative decoupled case.  This indicates that for the random input frame described with $\alpha$ given by equation \ref{eq:functional_alpha} the average magnitude difference between the truth image and the image decoupled using non-linear iterative method is linear with read noise scaling from 0 to 5 values as the read noise RMS goes from 0 to 100 values.}
\label{fig:3.2.4}
\end{figure}

However, it is known that the LR  deconvolution technique is designed to be noise resistant\cite{Lucy74}\cite{Richardson72}.  To explore this, cases where read noise scales from 0 RMS to 100 RMS were explored. As expected and illustrated in figure\ref{fig:3.2.5}, the IPC present case has no scaling mean absolute error.  The iterative decoupling's mean absolute error scales linearly with RMS read noise as shown in figure\ref{fig:3.2.4}.  The LR deconvolution's mean absolute error scales with RMS error of the read noise distribution in a sub-linear fashion converging to linear at high RMS, as seen in figure\ref{fig:3.2.5}.

\begin{figure}
\begin{center}
\begin{tabular}{c}
\includegraphics[height=6.0cm]{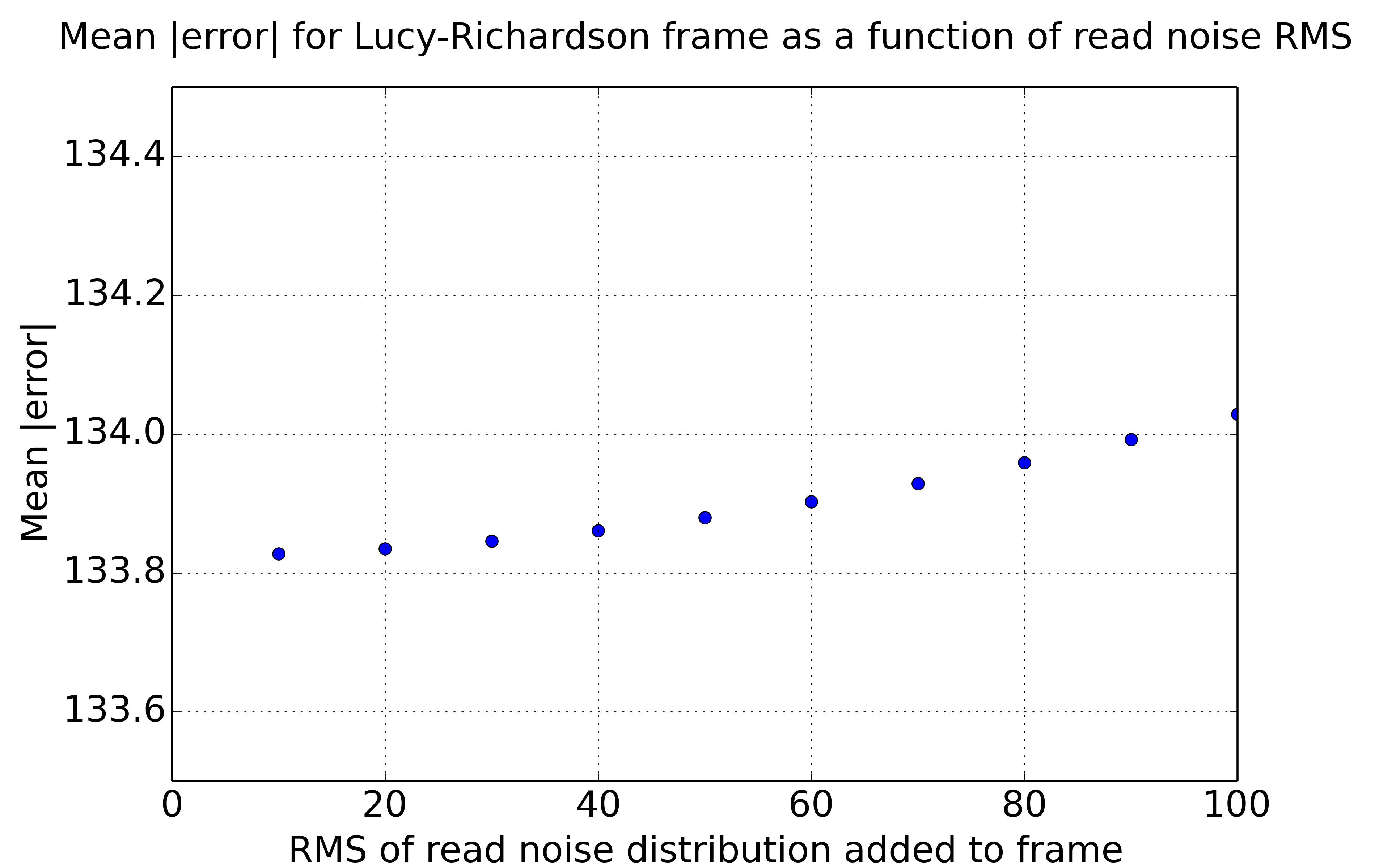}
\end{tabular}
\end{center}
\caption{Mean absolute error as a function of read noise for the LR decoupled case.  This indicates that for the random input frame described with $\alpha$ gvein by equation \ref{eq:functional_alpha} the average magnitude difference between the truth image and the image decoupled using non-linear iterative method scales non-uniformly with read noise, scaling from 133.8 to 134.0 as the read noise RMS goes from 0 to 100 values.}
\label{fig:3.2.6}
\end{figure}

It can be seen through comparison that though the growth behavior of the mean absolute error is slower in the case of LR deconvolution, the mean absolute error of the non-linear decoupling still allows for more accurate correction even with 100 RMS read noise distributions.  In fact, for the LR deconvolution to perform better, the read noise distribution introduced would have to be on the order of 800 RMS error.  Both on average and when evaluated pixel by pixel, the non-linear decoupling is more effective at removing IPC unless read noise is exceedingly large (on the order of 1\% of the array's saturation).

\subsection{PSF}

\begin{figure}
\begin{center}
\begin{tabular}{c}
\includegraphics[height=6.0cm]{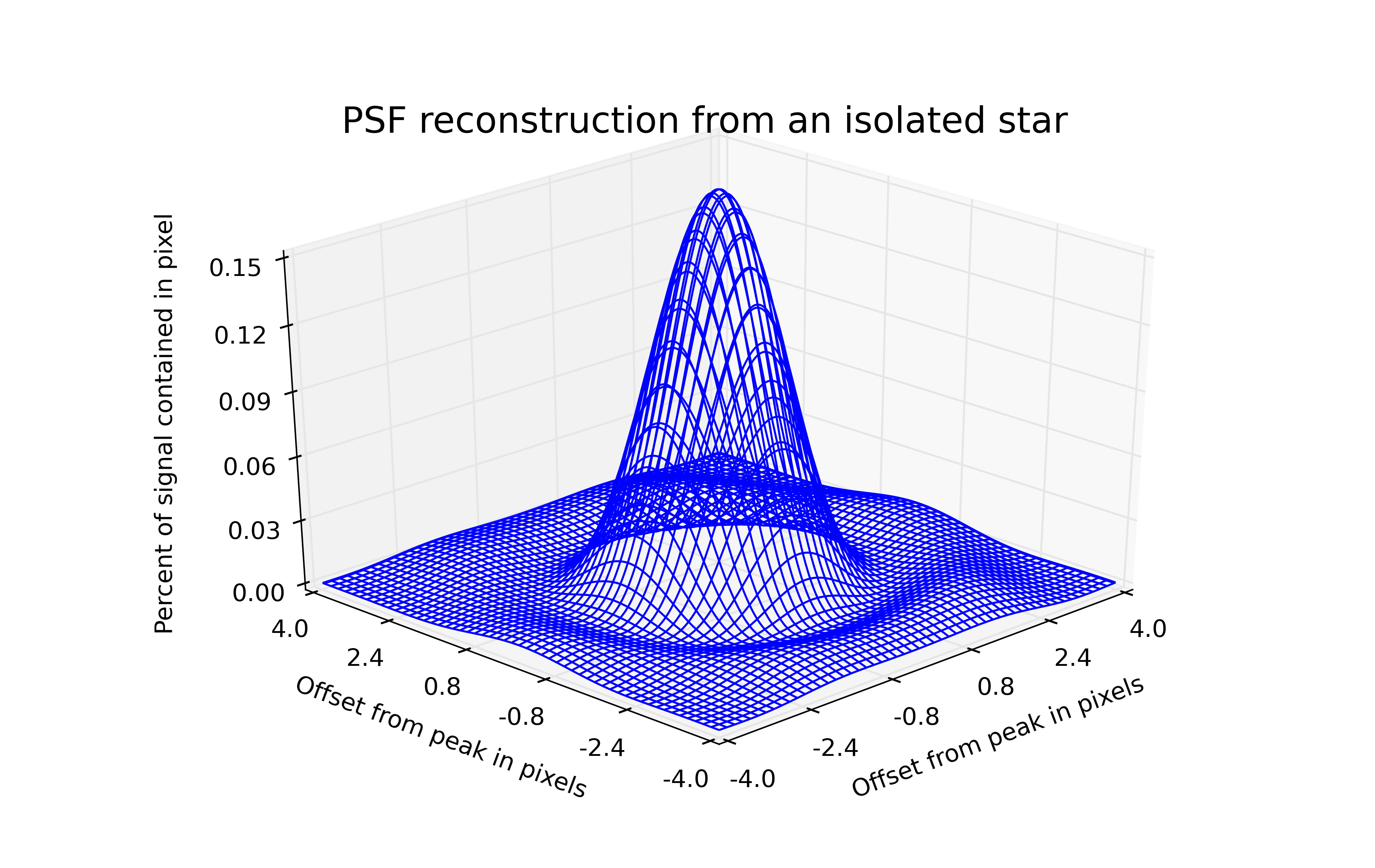}
\end{tabular}
\end{center}
\caption{3-dimensional reconstruction of the PSF in the non-linear iterative decoupled case.  This representation is used for PSF fitting.}
\label{fig:3.3.1}
\end{figure}

In this section we will examine the impact that IPC has on the structure of PSFs as the incident signal varies.  We will prescribe a particular coupling coefficient which matches the form presented previously in equation\ref{eq:functional_alpha}.  There is a distinction of kind between the blurring due to a diffraction and the blur from IPC;  diffraction is continuous whereas IPC blur occurs discretely between pixels.  The results presented here appear continuous as they are three dimensional cubic spline interpolations from data generated with a one quarter pixel movement in the x and y directions.  A sample three dimensional PSF result is shown in figure\ref{fig:3.3.1}.  By using an interpolation method, the PSF can be treated as a continuous mathematical object rather than just the discrete sampling.  This allows for better visualization of the PSF's distortion as well as greater ease in application for PSF fitting techniques as will be outlined in a later section of this work.

To obtain these PSF values the same data processing chart as presented earlier is used on a point source with the following exceptions and stipulations:  
\begin{enumerate}
	\item The background level is set to zero.  
	\item There is no Poisson sampling of the flux map.  
	\item The read noise values are all set to zero: zero mean and zero variance.  
	\item The values are normalized to unit volume.
\end{enumerate}
Three point source intensities were examined using this technique.  From here out they will be referred to as 'bri	ght', 'mid' and 'dim'.  The bright point source had an intensity such that with the WebbPSF F405N filter\cite{Perrin14}, post diffraction it would nearly saturate the central pixel at a 10ks exposure time.  This corresponds to a source of magnitude $\approx$ 21.4.  The dim source corresponds approximately to the sensitivity limit given through this filter with a 10ks exposure\cite{Beichman12} (i.e. a 25.3 magnitude source).  The mid source is a healthy midpoint between the two corresponding to the same 10ks exposure time observing a source of magnitude $\approx$ 23.1.

\begin{figure}
\begin{center}
\begin{tabular}{c}
\includegraphics[height=6.0cm]{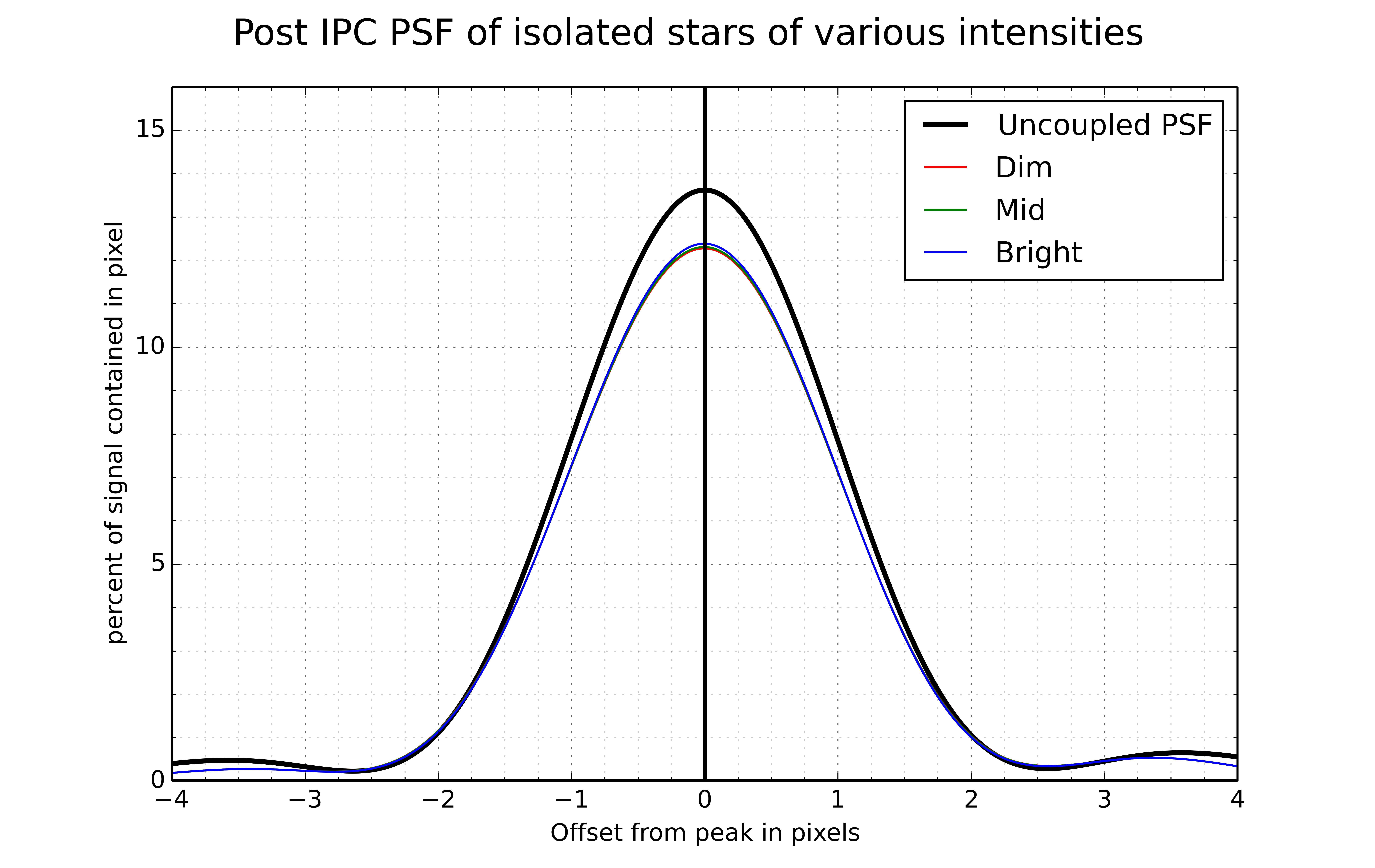}
\end{tabular}
\end{center}
\caption{PSF cross section through the peak for each intensity PSF.  the uncoupled PSF line represents both the true image and the decoupled image, as both are identical in this case.  Note that as a cross section, this function does not have unit area.}
\label{fig:3.3.2}
\end{figure}

\begin{figure}
\begin{center}
\begin{tabular}{c}
\includegraphics[height=6.0cm]{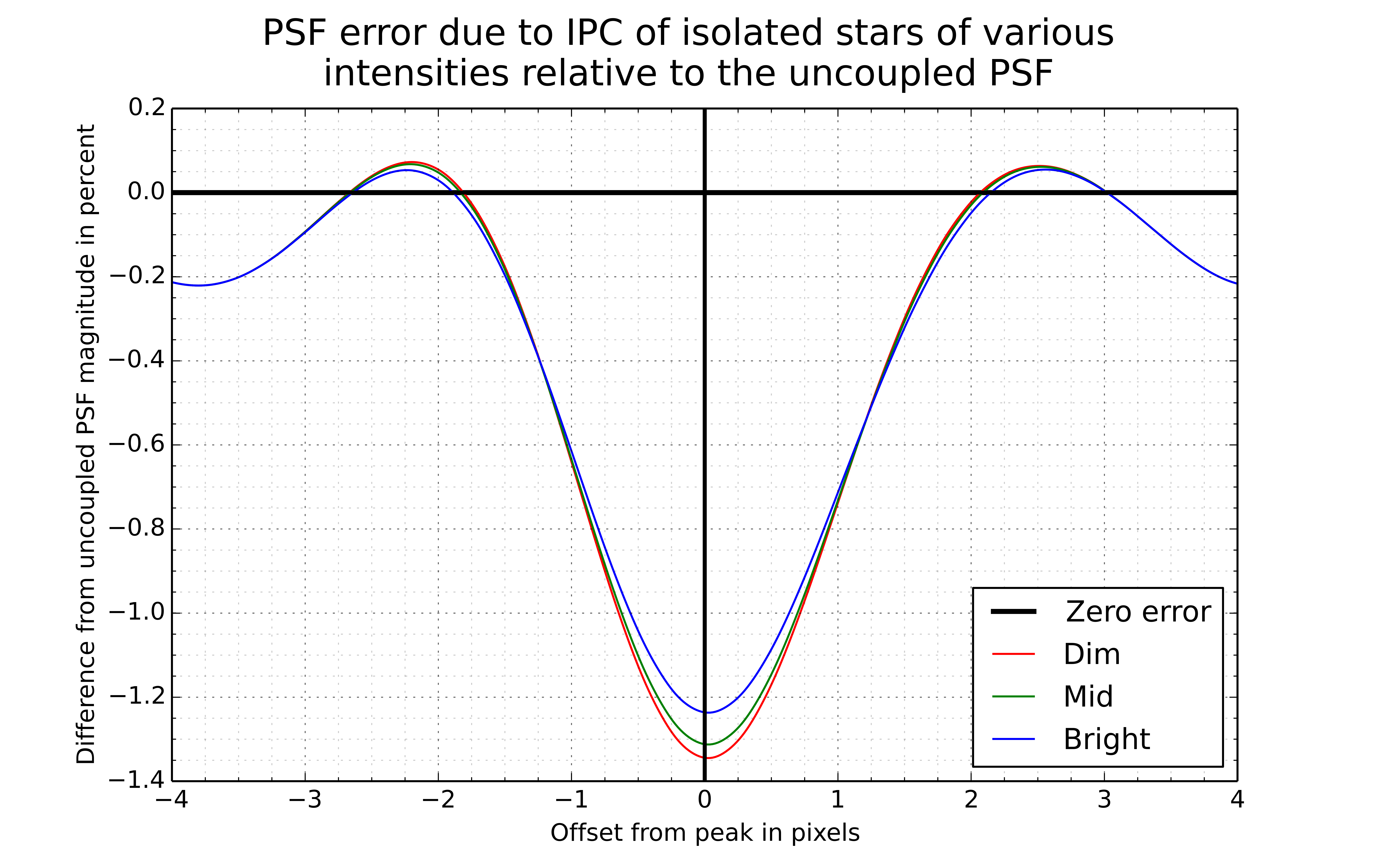}
\end{tabular}
\end{center}
\caption{Difference between true PSF and the IPC coupled PSF for each intensity.  This shows that IPC causes the PSFs to have a lower peak but also contains regions where the PSF is over-estimated.  IPC causes energy to shift away from local maxima and towards local minima.}
\label{fig:3.3.3}
\end{figure}

A common parameter for describing the shape of a PSF is the full width half max (FWHM).  The signal dependence of $\alpha$ results in stars of differing brightness having different FWHMs. These changes are described in the table\ref{table:FWHM}.  This result indicates that more fair objects will appear to have greater spacial extent than their higher flux counterparts.
\begin{center}
\begin{tabular}{|m{2.5cm}|m{2.5cm}|m{4.0cm}|}
\cellcolor{blue!33} PSF &  \cellcolor{blue!33}FWHM[pix] & \cellcolor{blue!33}FWHM [arcsec] \\
\hline
\hline
WebbPSF\cite{Perrin14} & 2.080 & 0.1352" \\
\hline
Dim & 2.186 & 0.1421" \\
\hline
Mid & 2.180 & 0.1417" \\
\hline
Bright & 2.169 & 0.1410"  \\
\hline
\end{tabular}
\end{center}
\begin{table}[H]
\caption{A table illustrating the widening of PSFs in the presence of IPC as the signal generated by the star drecreases as simulated for the JWST NIRCam long wavelength channel through F405N.}
\label{table:FWHM}
\end{table}

Both the true and decoupled frames after these operations are identical to each other across point source intensity within numerical error.  The IPC present frames in these cases are distinct from the true and decoupled frames as well as distinct from each other.  Due to the functional nature of the IPC applied, the more intense the point source the taller, narrower, and closer to true, the PSF is.  Equivalently, the lower flux levels give rise to a shorter, fatter, and more distorted PSF as seen in figure\ref{fig:3.3.2}.  The error introduced from IPC causes a decrease in the peak brightness of the PSF on the order of 1.2 to 1.4\%.  Additionally, the difference in PSF peak of a bright star relative to a dim star is on the order of 0.15\% as can be seen in the error cross section presented in figure \ref{fig:3.3.3}.

However these PSFs do not tell the full story of IPC coupling.  In a crowded field, when the post diffraction flux map of a scene results in overlap between sources, the IPC coupling experienced will be different than of a point source in isolation.  The simple case considered here is a system composed of two point sources separated by some distance.  In trends, IPC pulls collected signal away from local maxima and into local minima.  When point sources are near each other, the signal generated from each star couples differently towards the neighboring star compared to away from it.

\section{Photometry and Astrometry using DAOphot}
With established PSFs and a method to simulate IPC on a given frame, we can now examine how, and to what extent IPC will impact particular measurements.  Here we will look at the impact that a signal dependent IPC can have on will have on measurements of flux and separation estimates for a binary star.  PSF fitting will be performed by using established star best fitting techniques. The python implementation of IRAF starfinder\cite{Tody93} and DAO photometry\cite{Stetson87} developed in the photutils library\cite{Bradley16} are used.  These algorithms take in the arguments of an image frame for processing and a series of frames containing a super-sampled PSF. The output is a best-fit location and integrated magnitude of each star.

In the frames, the separation between the stars is varied from 2.0 pixels to 5.0 pixels, with an additional sample with separation of 20.0 pixels.  One star was given a brightness level called 'bright', 'mid' or 'dim' corresponding to the definitions given earlier.  The second star was given a brightness as a fraction of the first star ranging from equal intensity to $2^{-3}$ the intensity in powers of two.  The first star's location was initially set ranging from centered on the pixel to directly on the edge in 0.25 pixel intervals in both $x$ and $y$, giving 16 starting configurations per star pair.  Additionally both a noise present and noise absent case were examined.  The noise absent case was designed to show the the theoretical ideal and the noise present case gives a realistic simulation of expected results.  In order to evaluate the effectiveness of the decouple, results were compared between the 'true' frames, the 'coupled' frames, and the 'decoupled' frames.  An emblematic subset of these results are presented here:

\begin{figure}
\begin{center}
\begin{tabular}{c}
\includegraphics[height=8.0cm]{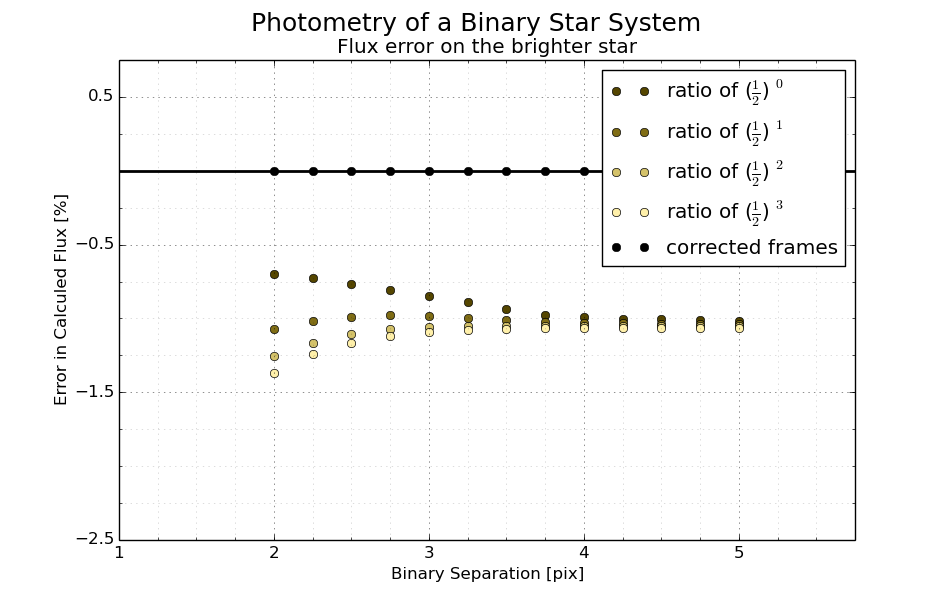}
\end{tabular}
\end{center}
\caption{Error introduced in flux estimation of the brighter star in an asymmetric binary system by IPC using DAOphot for fitting of the F405N WebbPSF to the IPC coupled data in the 'bright' star case.}
\label{fig:4.0.1}
\end{figure}

\begin{figure}
\begin{center}
\begin{tabular}{c}
\includegraphics[height=8.0cm]{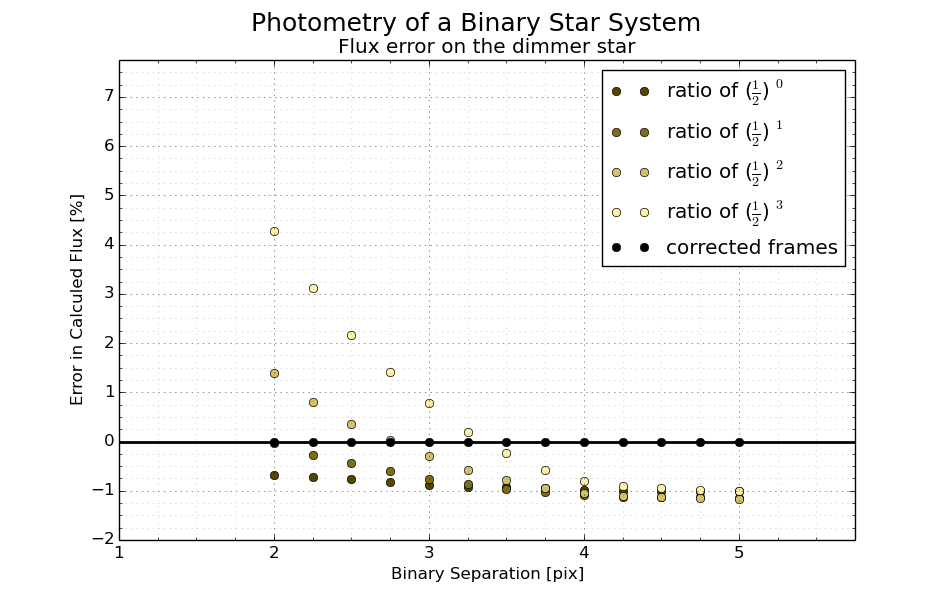}
\end{tabular}
\end{center}
\caption{Error introduced in flux estimation of the dimmer star in an asymmetric binary system by IPC using DAOphot for fitting of the F405N WebbPSF to the IPC coupled data in the 'bright' star case.}
\label{fig:4.0.2}
\end{figure}

\begin{figure}
\begin{center}
\begin{tabular}{c}
\includegraphics[height=8.0cm]{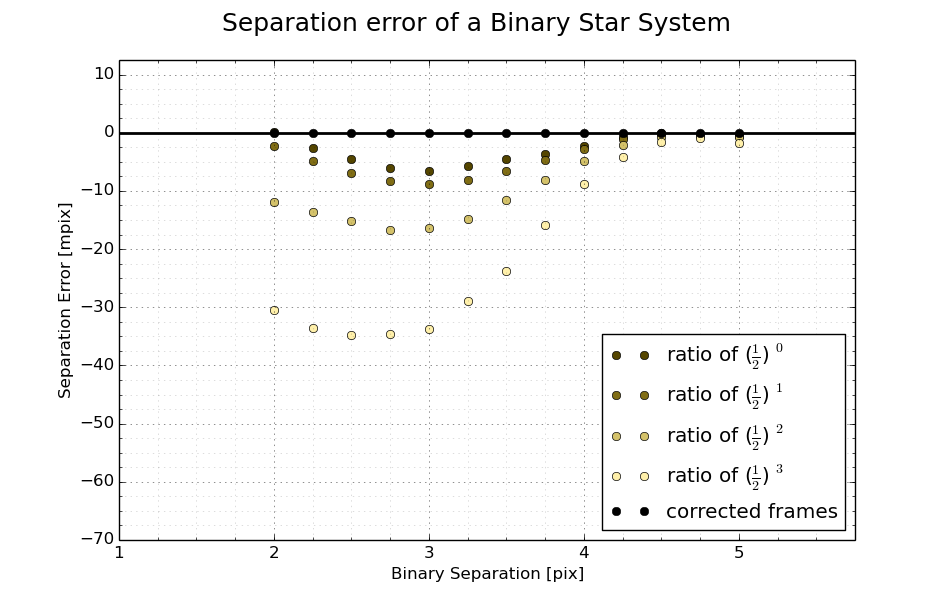}
\end{tabular}
\end{center}
\caption{Error introduced in separation estimation of an asymmetric binary system by IPC using DAOphot for fitting of the F405N WebbPSF to the IPC coupled data in the 'bright' star case.}
\label{fig:4.0.3}
\end{figure}

It can be seen from analysis of these frames that if ignored, an IPC on the order of 1\% to nearest neighbor can cause an error in accuracy of photometric estimates on the order of a few percent as seen in figures\ref{fig:4.0.1} and\ref{fig:4.0.2} or worse when attempting to discern a dim object from a bright neighbor.  This is due to the way in which the PSF distorting in the presence of IPC causing power that should be attributed to the brighter source, to instead be attributed to the more dim source.  Because there is a region of overlap between the two stars, they no longer blur independently.  The output from the sensor after IPC coupling occurs is not the sum of two coupled PSFs;  it is instead the coupling of the sum of two PSFs.  Because IPC coupling is not a commutitive operation, this distinction results in a break down of an assumption made by PSF fitting techniques.  The image as a whole can no longer be represented as a linear combination of stars with the same PSF.  Flux incident from each star is coupled more strongly towards the center of the binary than away from it.  This results in estimates of the center to center distances can being inaccurate on the order of tens of millipixels as seen in figure\ref{fig:4.0.3}.  If properly corrected for, the error in flux can be dropped to the level of hundredths or thousandths of a percent and the error in separation, dropped to the level micro-pixels.  Further figures examining the full range of parameters explored are available at \url{https://github.com/Donlok/Photometry_Astrometry}.
  
From equation\ref{eq:functional_alpha} as informed by simulations\cite{Donlon17} and previous observations\cite{Cheng09}, the greatest fractional coupling occurs when the signal difference between adjacent pixels is smallest.  In the case of confused point sources, this signal difference is larger on the exterior side than on the interior side.  This results in the interior side of the PSF of each star experiencing a greater coupling than the exterior side, causing energy to appear pulled towards the center.  Additionally, a higher fractional coupling occurs a the signal strength in any single pixel  decreases.  As a result, the PSF distortion is the most severe on confused systems where the brightness of the star is the lowest.  IPC's most severe impact on astrometric and photometric accuracy will occur when examining objects which are nearest to the sensitivity limit of the imaging system.

\section{Conclusion}
\begin{figure}
\begin{center}
\begin{tabular}{c}
\includegraphics[height=6.0cm]{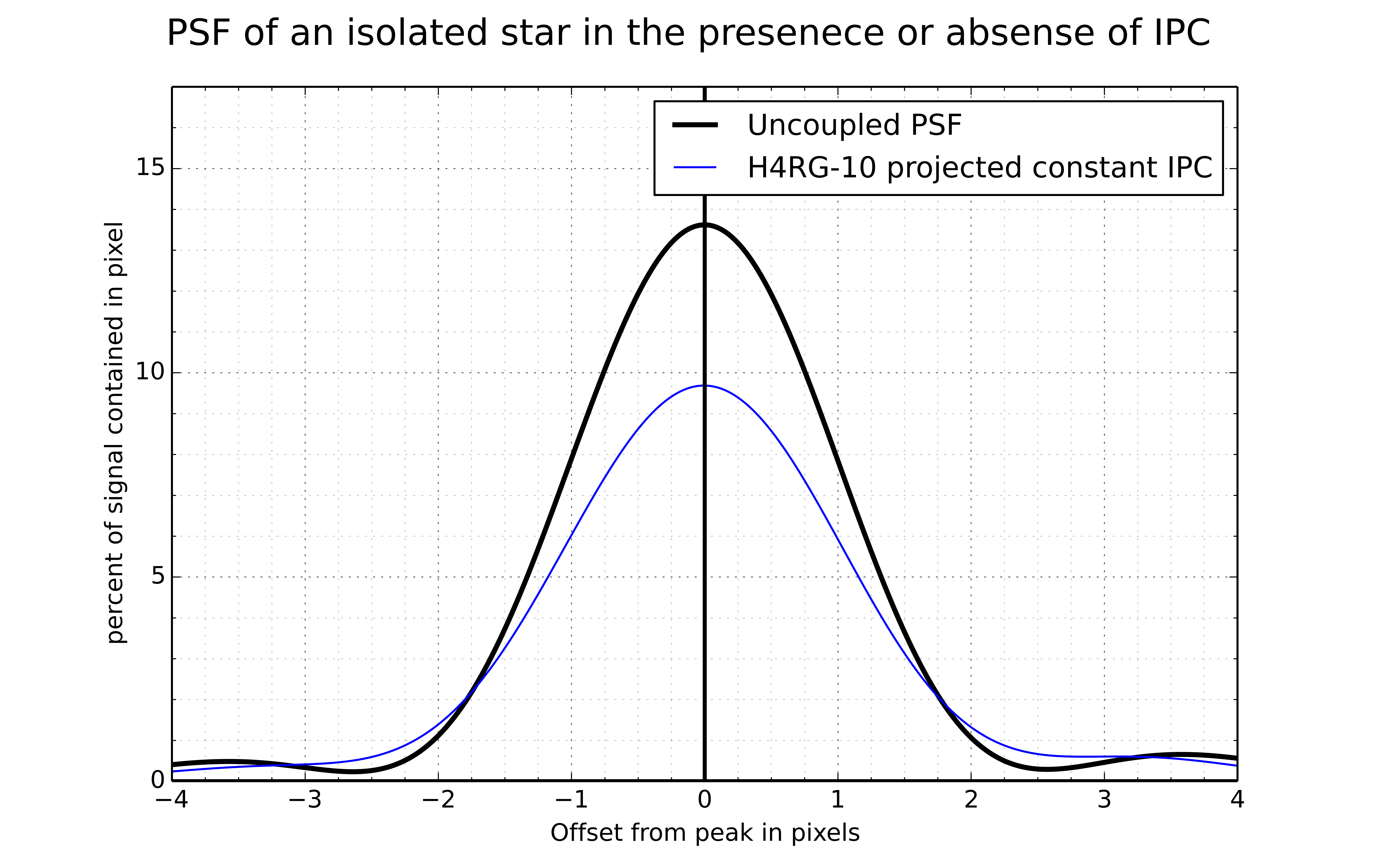}
\end{tabular}
\end{center}
\caption{PSF cross section using F405N filter for the diffraction pattern and a constant IPC kernel obtained by using a single pixel reset characterization method presented by Dudik, et al.\cite{Dudik12}.}
\label{fig:5.0.1}
\end{figure}

The non-linear iterative decoupling algorithm presented here is capable of completely removing the impact of IPC in the mathematically abstract case.  In the applied case, where read noise is present,  it is capable of signal restoration with error proportional to the read noise magnitude.  It requires a well characterized IPC, but, unlike other methods, it can account for an IPC which varies with signal level.  Failure to correct, or improper correction of, IPC can result in the introduction of systematic errors into the data which can result in incorrect scientific conclusions.  As missions in the infrared begin to transition from arrays using the H2RG readout circuitry to the smaller pixels of the H4RG-10 circuitry, IPC will continue to become larger and present a more significant problem.  Some measurements for H4RG-10 HgCdTe arrays indicated $\alpha$ on the order of 8 \% resulting coupling out of the central pixel on the order of 35\%\cite{Dudik12}.  If correction of IPC is not performed, this could result in erroneous conclusions and unnecessary imprecision from missions such as WFIRST.  In the case of higher couplings, diagonal and second-neighbor couplings can rise to non-negligible levels\cite{Dudik12}.  Figure \ref{fig:5.0.1} shows the PSF degradation that would be caused to the F405N WebbPSF in the presence of a static $\alpha=8\%$.  The FWHM would be expected to broaden from 2.080 pixels to 2.249 pixels.  This is an 8\% increase in FWHM prior to introduction of any signal dependence.  The greater IPC in the next generation of sensor will result in further image degradation and will require careful consideration regarding characterization, and correction.

The decoupling technique described here in equation\ref{eq:full_dec} for nearest neighbor coupling is easily extensible to coupling kernels of any size through expansion of the indexing of the sums.  In fact, as presented, equation\ref{eq:full_dec} includes diagonal coupling, though it has been set to zero in the implementations and examples provided.

\acknowledgments 
The research described in this paper was done by the authors with financial support by NASA through contract NAS5-02105. This research is based on simulations run on the BlueHive2 computing cluster supported by the University of Rochester Center for Integrated Computing with a faculty affiliation through Craig McMurtry. We thank both Craig McMurtry and the University of Rochester Research Computing staff for their assistance.


\bibliography{./report.bib} 
\bibliographystyle{ieeetr}

\end{spacing}

\end{document}